\documentclass[preprint]{aastex62}

\graphicspath{{./}{figures/}}

\usepackage{xcolor}

\let\a=\alpha    
    
 \let\t=\tau 
\let\r=\rho
\let\vp=\varphi 
	    \let\D=\Delta  
 
\let \x = \chi

\def\be{\begin{eqnarray}}
\def\ee{\end{eqnarray}}
\def\ba{\begin{array}}
\def\ea{\end{array}}

\def\td{\tilde}

\def \Msun {M_{\odot}}

\def \yr {\mathrm{yr}}

\def \kms {\mathrm{km \, s^{-1}}}
\def \cm {\mathrm{cm}}

\def \earth {\oplus}

\shorttitle{S0-2, G1, G2 and Sgr A*'s flaring activity of 2019}
\shortauthors{Lena Murchikova}

\begin{document}

\title{S0-2 star, G1- and G2-objects and flaring activity \\of the Milky Way's Galactic Center black hole in 2019}

\correspondingauthor{Lena (Elena M.) Murchikova}
\email{lena@ias.edu}

\author[0000-0001-8986-5403]{Lena Murchikova}
\affiliation{Institute for Advanced Study \\
1 Einstein Drive \\
Princeton, NJ 08540, USA}

\begin{abstract}

In 2019, the Galactic center black hole Sgr A* produced an unusually high number of bright near-infrared flares, including the brightest-ever detected flare \citep{2019ApJ...882L..27D,Gravity2020}. 
We propose that this activity was triggered by the near simultaneous infall of material shed by G1 and G2 objects due to their interaction with the background accretion flow. We discuss mechanisms by which S-stars and G-objects shed material, and estimate both the quantity of material and the infall time to reach the black hole.

\end{abstract}



\section{Introduction} \label{sec:intro}

There is a black hole at the center of our Galaxy. Its electromagnetic counterpart -- Sgr A* -- is a non-thermal and variable source. Its overal spectral shape is dominated by synchrotron emission with a cooling break.
Sgr A* shows flaring activity ranging from minutes to hours in radio, millimeter (mm), sub-millimeter (submm), near infrared (NIR), and X-ray frequency ranges. Here we discuss the NIR flaring activity, which is particularly well studied due to regular monitoring of stellar orbits around the Galactic Center in NIR \citep{Genzel10, Morris12, Witzel2018, Gravity2020}. 


There is no consensus about the precise mechanism of the flares and about their physical origin. They may be caused by a population of non-thermal electrons \citep{Dodds-Eden2010,2017MNRAS.468.2447P}, magnetic reconnection events \citep{2012AJ....144....1Y,2003ApJ...598..301Y}, or a short term increase of the accretion flow rate into the black hole at the event horizon. 

In May 2019, \citet{2019ApJ...882L..27D} [hereafter, Do19] observed an unprecedented bright NIR flare. The flare lasted $\sim 2.5$ hours. During this time the flux of Sgr A*
reached 6 mJy, an increase of $>100$ times its typical quiescent state. It was also at least twice as bright as the brightest flare previously observed. The flare was caught when the flux was already decreasing, therefore the values above are lower limits for the duration and the brightness of the flare. The activity characterized by an unusual number of strong flares continued until at least the end of 2019 \citep{2019ApJ...882L..27D, Gravity2020}. (At the time this work is written no data beyond 2019 is available.) Throughout this paper we call this ``the flaring activity of 2019'', for short. 

Do19 suggested that this unusual number of bright flares may be due to the physical changes in the accretion activity of Sgr A* possibly caused by the passage of the S0-2 (S2) star \citep{2019Sci...365..664D,Gravity2018} or the G2-object \citep{2017ApJ...847...80W,Gillessen19drag, Eckart2013}.
In this work we explore whether Sgr A*'s flaring activity of 2019  could have been caused by a temporary increase of accretion rate onto the black hole due to infall of the material left after the passage of G-objects or stars.
We use the timestamp of 2019.36 (corresponding to the extraordinary flare reported by Do19) to derive various delay times between the flaring activity of 2019 and the close flybys.

The paper is organized as follows. In Section \ref{sec:origin} we discuss the mechanisms by which the stars and the G-objects shed material along their orbits and estimate the amount of leftover material. In Section \ref{sec:time}, we estimate the amount of time required for the leftover material to reach the black hole. In Section \ref{sec:disc},
we compare the calculated physical infall time with the observed delay time between the pericenter passages of the S0-2 star and G-objects and the flaring activity of 2019. We then use this comparison to identify which of the recent close passages of G-objects and stars could have caused the flaring and present a schematic description of the process. Through the text we use au as a unit of distance. For conversion from arcsec to au we use the distance to Sgr A* as $8$ kpc. The Schwarzschild radius of a $4 \times 10^6 \Msun$ black hole, like Sgr A*, is $0.1$ au.

\section{The leftover material}\label{sec:origin}

We first consider the star making the closest approach to Sgr A* -- S0-2 star -- and its winds (see also \citealt{Loeb2004}). S0-2 is a $\sim 14 \Msun$ star of type B0-B2.5. A conservative upper limit on the S0-2's mass-loss rate is $< 3 \times 10^{-7} \Msun \, \yr^{-1}$ \citep{Martins08}. The limit is based on a non-detection of Br$\gamma$ ($\mathrm{n}=7 \to 4$) emission from the mass loss material. At larges mass loss rates the Br$\gamma$ emission would be filling the star's Br$\gamma$ absorption profile. The estimate actually constrains the density of the mass loss material $\r \sim \dot{M}/v_\infty,$ where $v_\infty$ is the wind velocity at infinity. \citet{Martins08} set $v_\infty=1000 \, \kms$ to obtain the quoted limit.

During 4 month of the S0-2's closest flyby of Sgr A* (2 months preceding and 2 months following the pericenter passage) its trajectory lies within about twice its pericenter distance from the black hole -- between $\sim 200$ au and 120 au -- and it would deposit $< 10^{-7} \Msun \simeq 0.03 M_\earth$ of material. Part of this material infalls into the black hole from the distances on average similar to the distances where the object which shed this material passed by while the rest escapes to much larger distances due to being launched from an object moving with an average velocity of  $\sim 7500 \, \kms.$ 

G-objects were discovered in NIR observations and their exact nature and origin are still debated (\citet{Ciurlo20, Stephan2019} and references therein). They have gaseous outer parts and likely have an embedded central object. For our purposes, the existence of the embedded object is not important, because it does not lead to the G-object to be gravitationally bound at the distances we are interested in, and because we are concerned only with the shredding of the gaseous cloud-like component as the object propagates through the background medium of the Galactic Center.

The G-objects shed material in a different way compared to the stars. While the S0-2 produces its wind at a constant rate, G-objects shed material due to the hydrodynamic friction. The density of the accretion flow increases the closer G-objects get to the black hole, and thus G-objects undergo the greatest hydrodynamic drag during their closest approach shedding most material there. It is also important that around the pericenter passage the G-objects are stretched, therefore it has higher contact area with the ambient medium and a lower density, both of which make shedding more effective. 

Let us consider the G2 first. We assume that G2's gaseous component has a mass of $\sim 3 M_\earth$ and radius $\sim 1.8 \times 10^{15} \cm$ \citep{Gillessen19drag}. This implies that G2's density is $\sim 4.5 \times 10^{5} \, \cm^{-3}.$ 
During $\sim 8$ month of the closest approach, the object stays within a couple of its periapse radii -- within 400 au -- and moves with average velocity is $V_{mean} \sim 6000 \kms.$
To estimate the amount of shed material we use the results of \citet{Forbes_Lin19}, who studied hydrodynamic shredding of clouds moving in a background medium $10^2$ times less dense than the cloud.\footnote{\citet{Forbes_Lin19} considered multiple clouds moving one after the other and studied the importance of the hydrodynamic shielding. However when the times are as short as $\sim 10 t_{cross}$ and the separation between the clouds is large the shielding is not important, and we can apply the results of the simulations to individual clouds shedding.} The 8 months of the closest approach correspond to $\sim 7 t_{cross},$ where $t_{cross}=r_{G2}/V_{mean},$ during which the G2 would lose $\sim 2 \%$ of its mass, or $\sim 2 \times 10^{-7} \Msun \simeq 0.06 M_\earth.$ Similarly to the case of the S0-2, only a part of the leftover material infalls into the black hole from the vicinity of the G2's pericenter, the rest escapes to much larger distances.

The properties of the G1 are less constrained than the properties of the G2 (see caption to Table \ref{tab:1} for the references to observations), but they are generally assumed to be similar. The G1 was followed from two years after the periapse. It is estimated that the G1 passed at $\sim 300-400$ au from Sgr A*. It stayed within about twice its periapse distance, i.e. within $\sim 600$ au, for 1.3 years or for $\sim 12 t_{cross}.$ Following \citet{Forbes_Lin19} we estimate that during the passage it lost $\sim 5\%$ of its mass -- $\sim 0.15 M_\earth \simeq 5 \times 10^{-7} \Msun$ -- due to hydrodynamic shredding. 
There are evidence that G1 could have deposited even more material: it was becoming fainter and smaller for the first several years of observations diminishing from 450 au in 2004 to less than 170 au in 2009 \citep{2017ApJ...847...80W,2015ApJ...798..111P}. Whether this diminishing started as the G1 approached close to the pericenter or it was due to other causes well after the passage, i.e. a collision with a cool $10^4$ K disk \citep{2019Natur.570...83M}, we may never know.

To put the these amounts of the leftover material deposited by the close flybys in context, we evaluate the amount of the background accretion flow material around Sgr A*.
We use the value of the accretion flow density at $\sim (2-8) \times 10^{3} \, \cm^{-3}$ at $100-200$ au obtained by \citet{Gillessen19drag} from the drag force on the G2. The simulations produce comparable densities (e.g. \citealt{2020MNRAS.492.3272R}).
The amount of the accretion flow material within $400$ au is then $\sim 3 \times 10^{-6} \, \Msun.$ The leftover material constitutes about 10\% of the background material, which is large enough to expect changes in the accretion flow rate, but small enough for the background material to be able to slow down the leftover material equalizing its orbital momentum with its own and allowing the infall to proceed as a part of the flow. 

Not all of the leftover material will be accreted onto Sgr A*. The further the material is deposited from the black hole the smaller the fraction of it that will be accreted. Large fraction of it will likely be blown away by outflows \citep{Blandford.Begelman99}. 

The material deposited by other S-stars will be further away from the black hole compared to the S0-2's leftover material.
Therefore, the ability of other S-stars to enhancing the accretion rate onto Sgr A* is diminished the further their pericenter and their trajectory from the black hole.
The stars are losing material at a constant rate during their passages, in contrast to G-objects that tend to deposit the bulk of it at the pericenter. Pericenter passages of S-stars are quite common and happen on average once a year (A. Ghez, private communication), which point at less routine events as a cause of the flaring activity of 2019. These arguments disfavor star passages as a cause of the Sgr A*'s flaring activity of 2019, compared to G-object's flybys.

\section{The infall time}\label{sec:time}

To calculate the infall time of the leftover material we assume that (i) the leftover material joins the accretion flow and is carried inward with it, and (ii) the initial distance for the infall is equal to the pericenter distance of the object depositing it.
The radial inflow velocity of the material can be approximately expressed as
\be
	v_R \simeq \a \left( \frac{H}{R} \right)^2 v_\varphi = \a \x \left( \frac{H}{R} \right)^2 v_K,
\ee
where $R$ is the radial coordinate,  $H$ is the scale height of the disk at the radius $R$, $\a$ is a dimensionless $\alpha$-viscosity parameter \citep{Shakura73}, and $\x = v_\vp/v_{K}$ parameterises deviation of the accretion flow's azumithal velocity ($v_\vp$) from the Keplerian velocity $v_K=\sqrt{{GM_{\mathrm{BH}}}/{R}}.$ 
The infall time from radius $R_{in}$ is
\be\label{eq:in.time}
	\t_{in}(R_{in}, \a \x) &\simeq& \int\limits^{R_{in}}_{0}	\frac{dR}{v_R(R)} = 
	\frac{2 R_{in}^{3/2}}{3\a \x \sqrt{GM_{\mathrm{BH}}}}. 
\ee
Here we neglected the effects of general relativity as $R_{in} \gg R_g,$ and set $H/R \sim 1$ as Sgr A* accretion is a thick disk accretion.

\section{Discussion}\label{sec:disc}

\begin{deluxetable}{cccccccccc}
\tabletypesize{}
\tablecaption{Recent notable close approaches to Sgr A*.} \label{tab:1}  
\tablewidth{0pt}
\tablehead{
\colhead{Obj. } & \colhead{Ref} & \colhead{Peri time} & \colhead{Peri dist, au} & \colhead{$t_{flare} - t_{peri}$} & \colhead{Infall time} & \colhead{Infall time} & \colhead{$t_{in}-\D T$} & \colhead{$\td{t}_{in}-\D T$}\\
\colhead{name } & \colhead{} & \colhead{$t_{peri}$} & \colhead{$R_{peri}$, au} & \colhead{$\D T,$ yr} & \colhead{$t_{in},$ yr} & \colhead{$\td{t}_{in},$ yr} & \colhead{yr} & \colhead{yr}
}
\startdata
S0-2&   & 2018.38$^{+<0.01}_{-<0.01}$ & 120$^{+<1}_{-<1}$ & 1.0$^{+<0.01}_{-<0.01}$     & 2.8$^{+<0.1}_{-<0.1}$  & 4.6$^{+<0.1}_{-<0.1}$ & 1.8$^{+<0.1}_{-<0.1}$ & 3.6$^{+<0.1}_{-<0.1}$  \\ 
G1  & P15 & 2001.6$^{+0.1}_{-0.1}$  & 417$^{+239}_{-239}$ & 17.8$^{+0.1}_{-0.1}$  & 18.0$^{+17.5}_{-13.0}$   & 29.9$^{+29.1}_{-21.6}$ & \bf{0.2}$^{+17.5}_{-13.0}$ & 12.1$^{+29.1}_{-21.6}$ 
\\ 
G1  & W17 & 2001.3$^{+0.4}_{-0.2}$  & 298$^{+32}_{-24}$ & 18.1$^{+0.2}_{-0.4}$  & 10.9$^{+1.8}_{-1.3}$ & 18.1$^{+3.0}_{-2.1}$ & -7.8$^{+1.8}_{-1.4}$ & \bf{0}$^{+3.0}_{-2.1}$
\\ 
G2  & W17 & 2014.2$^{+0.03}_{-0.05}$  & 201$^{+13}_{-13}$ & 5.2$^{+0.05}_{-0.03}$   & 6.0$^{+0.6}_{-0.6}$  & 10.0$^{+1.0}_{-1.0}$ & \bf{0.8}$^{+0.6}_{-0.6}$ & 4.8$^{+1.0}_{-1.0}$  \\ 
G2  & G19 & 2014.49$^{+0.06}_{-0.06}$ & 127.5$^{+2.5}_{-2.5}$ & 4.9$^{+0.06}_{-0.06}$   & 3.1$^{+0.1}_{-0.1}$  & 5.2$^{+0.15}_{-0.15}$ & -1.8$^{+0.1}_{-0.1}$ & \bf{0.3}$^{+0.2}_{-0.2}$ 
\\
\enddata
\tablecomments{Selected notable flybys by Sgr A*. The table lists each object's name, the time of the pericenter passage ($t_{peri}$), the distance to the pericenter ($R_{peri}$), the delay time ($\D T=t_{flare} - t_{peri}$), which is the time passed since the object's pericenter passage at the time of the extraordinary flare on 2019-05-13 ($t_{flare}$) used here as a time stamp for the flaring activity of 2019, the infall time for the object's leftover material to reach black hole calculated with eq. \ref{eq:in.time025} ($t_{in}$), and the infall times calculated with eq. \ref{eq:in.time2} ($\td{t}_{in}$). The last two columns list the differences between in the infall times and the delay times. When this value is close to zero (marked in bold), an association between the object's flyby and the flaring activity of 2019 may be established. The data for the S0-2 is from \citet{2019Sci...365..664D, Gravity2018}, for the G1 -- from  \citet{2015ApJ...798..111P, 2017ApJ...847...80W}, and for the G2 -- from \citet{2017ApJ...847...80W,Gillessen19drag}. For the ease of referencing, in those cases when different references provide substantially different estimates for the pericenter distances and the passage times we list the first letter of the titular author and the year of publication in the second column. The uncertainties listed are the observational uncertainties.
}
\end{deluxetable}

In Table \ref{tab:1} we list the periapce distances ($R_{in}$), the time of the pericenter passages ($t_{peri}$), and the delay times between the pericenter passage of each object and May 13, 2019, the timestamp for the flaring activity of 2019 ($\D T$) for the three most notable recent passers by Sgr A* -- the S0-2 star, G1-, and G2-objects. The G1 and G2 have two entries each corresponding to two independent determinations of their orbital parameters.

From the Table one can see that the G1 and G2 orbital parameters have an interesting property. The material left by the G1 and G2 during their close passage will reach the black hole nearly simultaneously and the time the material reaches the black hole coincides with the flaring activity of 2019. If we adopt $\a \x = 0.025,$ as in \citet{2019Natur.570...83M}, we obtain 
\be\label{eq:in.time025}
t_{in}(R_{in}) \simeq \t_{in}(R_{in}, \a \x=0.025)=2.11 \, \yr \, \left( \frac{R_{in}}{100 \mathrm{au}} \right)^{\frac{3}{2}}.
\ee
It follows that $t_{in} \simeq \D T$ for the G1 parameters determined by \citet{2015ApJ...798..111P} and the G2 parameters determined by \citet{2017ApJ...847...80W} (see bold in column 8).
If we adopt a slightly different $\a \x = 0.015:$ 
\be\label{eq:in.time2}
	\td{t}_{in}(R_{in}) = \t_{in}(R_{in}, \a \x=0.015)  \simeq 3.52 \, \yr \, \left( \frac{R_{in}}{100 \mathrm{au}} \right)^{\frac{3}{2}},
\ee
we find that $\td{t}_{in} \simeq \D T$ for the infall time for the  G1 parameters determined by \citet{2017ApJ...847...80W} and the G2 parameters determined by \citet{Gillessen19drag} (see bold in column 9). Both choices of $(\a \x)$-parameter are reasonable. Simulations suggest $\a \sim 0.1$ and the sub-Keplerian parameter was estimated by \citet{2019Natur.570...83M} at $\x \sim 1/4.$  

One might object that by adjusting the parameter $(\a \x),$ it is easy to obtain virtually any infall time. It is indeed possible to match the delay time and the infall time for any {\it{one}} given $R_{in}.$ Here however we have two independently determined infall times of two independent passage events matching their corresponding delay times. This cannot be obtained with any choice of single parameter $(\a \x)$ for randomly chosen delay times.

We conclude that the likely cause of the May 2019 flare reported by Do19 and the rest of the flaring activity of 2019 is the nearly coincident infall of the leftover material from the passages of the G1 and G2 objects and the continued infall of the tail-end of the shed G1 and G2 material. This temporary increases the mass accretion rate onto Sgr A*, stimulate more frequent production of reconnection events and/or shocks, and enable more energy dissipation, thus increase the probability of producing bright flares.

The following is a schematic of the process. During the passage of the G-objects by Sgr A*, part of their material is stripped by the background accretion flow material, resulting in hydrodynamic shredding. The stripping is particularly effective in the vicinity of their pericenter passage, where the density of the background material is the highest and the G-objects are tidally stretched, thus having a decreased inner density and increased contact surface area. The amount of the leftover material constitutes an approximate $10\%$ increase over the background material. Part of the stripped material escapes to much larger radial distances from Sgr A* before eventually slowing down. We disregard the contribution of this material: the further the material's initial infall radius, the smaller the fraction of it that reaches the black hole due to the expected presence of outflows. The other part of the leftover material efficiently loses its orbital momentum due to interaction with the background accretion flow, and slows down in the vicinity of the pericenter of the G-object. Eventually this material falls toward the  back hole. During the infall the material would fragment and spread, a large portion of it would be blown away by outflows. The part that would reach the black hole would reach it in the form of overdensities. Each would then temporarily increase an accretion flow onto Sgr A*. These overdensities are thus responsible for the increased number of strong flares in 2019 (and possibly later).

The influx of additional material must affect the populations of thermal and non-thermal electrons, causing increased mm flux. 
The absence of a rise of mm flux in 2019 would imply that the influx of additional material does not affect the population of electrons, and thus would rule out the population of non-thermal electrons as a cause of the NIR flares. In this case we would conclude that the NIR flares of Sgr A* are likely caused by magnetic reconnection. The observations of mm emission from Sgr A* during the flaring activity exist and we believe should be reported at some point.

The difference between the delay time and the infall times from the S0-2's pericenter is 1.8 years. This mismatch leads us to reject the S0-2 flyby as a cause of the flaring activity of 2019. However since the star is massive, possibly with strong winds and passes the closest to the black hole it may cause some flaring at a later time. We estimate that it should take the S0-2's leftover material about 2.8 years (assuming $\a \x = 0.025$) to reach Sgr A* from the time of the pericenter passage. Thus we should expect moderate increased flaring activity around 2021.2, i.e around March 2021. Alternatively, if the $\a \x = 0.015$ parameters are better for describing the properties of the infall, we expect moderate flaring activity 4.6 years after the pericenter passage, i.e. around 2023.0 or January 2023. Observation of increased flaring activity around either of these dates would allow us to constrain the accretion flow parameters to within $\sim 10^3 R_{s},$ as well as to confirm (or reject) that the S0-2 star can cause elevated flaring activity.

One may speculate that the timing of the G1 and the G2 approaches is a coincidence and that the increased flaring activity of 2019 is unrelated to any flybys. The accretion rate onto Sgr A* is expected to be variable with time \citep{2020MNRAS.492.3272R} on the scale of tens of years. So perhaps the increased activity is just the result of this long term variability. Indeed we cannot rule out this possibility at the moment, but we will within next $\sim 1-2$ years. If the flaring activity of 2019 is the result of the G-objects flybys and of the infall of the material they shed, the flaring activity will cease within the time required for the bulk of the G1's and the G2's leftover material to reach the black hole. If the flaring activity is due to long term accretion flow variations, it will persist for years to come.


Extensive simulations of the infall of the G2 related material and its influence on Sgr A* luminosity were conducted nearly a decade ago, see e.g. \cite{Moscibrodzka2012,Sadowski2013}. Most studies were performed in the context of violent tidal disruption and infall of G2, then considered a cloud, onto the black hole. This did not occur. Over the recent years our knowledge of the G1 and G2 trajectories have much improved as well as more accurate models of the accretion flow around the black hole became available. We hope that the reasoning presented here will inspire a suit of modern numerical studies of G1 and G2 interaction with the accretion flow, their influence on the black hole accretion and their ability to cause Sgr A*'s extraordinary flaring of 2019.

\section*{Acknowledgments}

I am grateful to Tuan Do whose visit to the IAS in autumn 2019 sprung my interest in the flaring activity of 2019 and who prompted me to conduct the original estimations, to Nadia Zakamska for encouragement and support with writing this paper, to Roger Blandford, Charles Gammie, Yuri Levin, Elias Most, Smadar Naoz, Sean Ressler, Scott Tremaine, and the anonymous referee for useful comments and suggestions.
LM's stipend at the Institute for Advanced Study is provided by the Corning Glass Works Foundation.






\bibliography{sgra_flares}{}
\bibliographystyle{aasjournal}




\end{document}